\documentclass[sigconf]{acmart}

\usepackage{booktabs} 
\usepackage{graphicx}
\usepackage{amsmath} 
\usepackage{color}

\graphicspath{{./figures/}}
\usepackage{epstopdf}
\DeclareGraphicsExtensions{.eps}

\usepackage{soul}

\usepackage{comment}
\usepackage{tabularx}
\usepackage{tabu}
\usepackage{pbox}
\usepackage{multirow}
\usepackage{url}
\usepackage{wasysym}
\usepackage{caption}
\usepackage{enumerate}
\usepackage{listings}
\usepackage[ruled, vlined]{algorithm2e}
\setcopyright{acmcopyright}

\copyrightyear{2021}
\acmYear{2021}
\setcopyright{acmcopyright}\acmConference[SAC '21]{The 36th ACM/SIGAPP Symposium on Applied Computing}{March 22--26, 2021}{Republic of Korea}
\acmBooktitle{The 36th ACM/SIGAPP Symposium on Applied Computing (SAC '21),
March 22--26, 2021, Virtual Event, Republic of Korea}
\acmPrice{15.00}
\acmDOI{10.1145/3412841.3441970}
\acmISBN{978-1-4503-8104-8/21/03}

\begin{document}
\title{DeepiSign: Invisible Fragile Watermark to Protect the Integrity and Authenticity of CNN}

\author{Alsharif Abuadbba}
\email{sharif.abuadbba@data61.csiro.au}
\orcid{1234-5678-9012}
\affiliation{%
  \institution{Data61, CSIRO, Australia \\ Cyber Security CRC}
}
\author{Hyoungshick Kim}
\email{hyoung.kim@data61.csiro.au}
\affiliation{%
  \institution{Data61, CSIRO, Australia \\ Sungkyunkwan University, South Korea }}
\author{Surya Nepal}
\email{surya.nepal@data61.csiro.au}
\affiliation{%
  \institution{Data61, CSIRO, Australia \\ Cyber Security CRC}
}
\newcommand{\sharif}[1]{\textcolor{blue}{Sharif: #1}}

\begin{abstract}
Convolutional Neural Networks (CNNs) deployed in real-life applications such as autonomous vehicles have shown to be vulnerable to manipulation attacks, such as poisoning attacks and fine-tuning. Hence, it is essential to ensure the integrity and authenticity of CNNs because compromised models can produce incorrect outputs and behave maliciously. In this paper, we propose a self-contained tamper-proofing method, called DeepiSign, to ensure the integrity and authenticity of CNN models against such manipulation attacks. DeepiSign applies the idea of \emph{fragile invisible watermarking} to securely embed a secret and its hash value into a CNN model. To verify the integrity and authenticity of the model, we retrieve the secret from the model, compute the hash value of the secret, and compare it with the embedded hash value. To minimize the effects of the embedded secret on the CNN model, we use a wavelet-based technique to transform weights into the frequency domain and embed the secret into less significant coefficients. Our theoretical analysis shows that DeepiSign can hide up to 1KB secret in each layer with minimal loss of the model's accuracy. To evaluate the security and performance of DeepiSign, we performed experiments on four pre-trained models (ResNet18, VGG16, AlexNet, and MobileNet) using three datasets (MNIST, CIFAR-10, and Imagenet) against three types of manipulation attacks (targeted input poisoning, output poisoning, and fine-tuning). The results demonstrate that DeepiSign is verifiable without degrading the classification accuracy, and robust against representative CNN manipulation attacks.
\end{abstract}

%
%



\maketitle

\section{Introduction}\label{sec:intro}

Convolutional Neural Networks (CNNs) are now popularly deployed in real-life applications such as autonomous vehicles \cite{intro:chen2015deepdriving} and drones \cite{intro:drones2019}. However, recent studies \cite{attacks:gu2017badnets,rw:adi2018turning} demonstrated that CNNs are inherently vulnerable to manipulation attacks, such as poisoning attacks \cite{rw:liu2017trojaning,rw:chen2017targeted} and fine-tuning \cite{rw:rouhani2018deepsigns}. In those attacks, the adversary can retrain the original CNN using some (intentionally crafted) samples with improper labels--such samples can be generated with a note marker, called \emph{backdoor}, which can be used as a trigger to activate the attack. Liu et al. \cite{rw:liu2017trojaning} further improved manipulation attacks by tampering the weights at hidden layers to secretly change the CNN model's behavior. Gu et al. \cite{attacks:gu2017badnets} demonstrated the feasibility of these attacks in a real-life autonomous vehicle application. For example, given a CNN model, the adversary retrains the CNN model with \emph{`stop sign'} images containing a backdoor so that \emph{`stop sign'} images with a backdoor would be incorrectly recognized as \emph{`speed sign'}. Therefore, it is essential to ensure the integrity and authenticity of CNN models against such backdoor attacks after deployment.  


Perhaps, a possible solution is to use cryptographic primitives (e.g., digital signature) to provide the integrity and authenticity of CNN models. In this case, however, the signature is additionally distributed and should also be securely protected. If the signature can be accidentally lost or intentionally removed in an attempt to cheat, how can we determine whether the CNN model is compromised? Perhaps, it is necessary to protect the presence of the signature itself. Furthermore, whenever a new CNN model is introduced, the model's signature is additionally needed. Therefore, we need to hold multiple signatures for multiple CNN models. This requirement would be burdensome for some computing environments that cannot provide secure storage holding such signatures. 

To overcome the limitation of cryptographic primitives requiring the protection of external and independent signature, we introduce a novel self-contained tamper-proofing method called DeepiSign as an alternative and complementary method to conventional cryptographic solutions. Our goal is to securely bind the model's signature to the model by embedding it in the model itself so that attackers cannot easily damage or remove the signature. Self-contained solutions based on watermarking have been intensively studied in recent times  \cite{rw:adi2018turning,rw:uchida2017embedding,rw:nagai2018digital,rw:zhang2018protecting,rw:merrer2017adversarial,rw:rouhani2018deepsigns}. However, such approaches have proven to be successful in asserting ownership but failed to protect the integrity in the face of security threats such as poisoning attacks. The existing watermarking models are designed to resist (stay unchanged) the modification by an adversary who wants to steal the CNN and claim its ownership \cite{rw:rouhani2018deepsigns}. To address the shortcomings of existing approaches, we are dedicated to investigating the following research questions: 


\vspace{0.1cm}
\noindent\textit{\textbf{Can we embed a self-contained mechanism inside a CNN model to ensure its authenticity and integrity by satisfying the following conditions: (C1) Minimal loss of the model accuracy; (C2) Ability to detect model manipulation attacks after deployment; and (C3) Sufficient security of the mechanism?}}

To answer these questions, we explore a widely used method in the multimedia domain, such as image, called \emph{invisible fragile watermarking}~\cite{stego:ched2010digital}. The sender of a message (Alice) hides the message into an image so that the (authorized) recipient (Bob) can only retrieve it, but the adversary (Eve) cannot tell whether a given image contains a message or not. Any change in the image renders the hidden secret invalid. Although fragile watermarking might be a promising solution as a self-contained method, its direct application to CNN models might violate the two following conditions: \textbf{(C1) Minimal loss of the model accuracy} and \textbf{(C3) Sufficient security of the mechanism itself}. The distortion due to the insertion of a watermark may not be a significant issue in the multimedia domain because small changes in multimedia contents could not be readily perceptible by the human eye (e.g., the presence of a few greyer pixels in an image is not a serious issue). However, in the context of CNN, it should be taken carefully due to the sensitivity of weights in the hidden layers, which might significantly impact the CNN model's performance. Also, it is challenging to protect the embedded watermark from attackers. If the embedded watermark is always located at fixed positions, the attacker can easily remove the embedded watermark. Therefore, it would be essential to locate the embedded watermark at dynamic positions randomly. 

In this paper, we propose a fragile watermark-based self-contained algorithm, called DeepiSign\footnote{\url{https://github.com/anonymousForNow/DeepiSign}}, that ensures both the integrity and authenticity of CNN models. DeepiSign is designed to satisfy the three conditions:

\textbf{(C1) To solve the accuracy degradation} issue inherent to fragile watermark distortion, we employ an algorithm based on a wavelet decomposition to transform weights at hidden layers in a CNN model from the spatial domain to the frequency domain. We apply appropriate derived scaling factors of $\delta$ and $\varrho$ to the identified less significant coefficients to minimize the impacts of distortions and preserve the model's accuracy. $\delta$ and $\varrho$ are experimentally derived numerical values to protect the sign and decimal precision. DeepiSign builds a unique secret related to each layer and utilizes detailed coefficients in the frequency domain to hide the secret and its hash value secretly inside the corresponding layer.

\textbf{(C2) To detect model manipulation attacks}, we provide extensive empirical evidences of the security of DeepiSign by performing several experiments on four pre-trained models (ResNet18 \cite{ds:resnet18:he2016}, VGG16 \cite{intro:simon2014VGG}, AlexNet \cite{intro:kriz2012alexnet} and MobileNetV2 \cite{sandler2018mobilenetv2}) using three datasets (MNIST \cite{ds:mnist:le1995}, CIFAR-10 \cite{ds:cifar10:kriz2009}, and Imagenet \cite{ds:deng2009imagenet}) against three types of manipulation attacks (targeted input poisoning \cite{attacks:gu2017badnets}, output poisoning \cite{rw:adi2018turning}, and fine-tuning \cite{rw:rouhani2018deepsigns}).
The experimental results show that DeepiSign is verifiable and secure without compromising the model accuracy.

\textbf{(C3) To ensure the security} of the watermark and make it undetectable, we randomize the location of the embedded watermark, which is determined by the initially shared parameters between the sender and the recipient. We perform mathematical steganalysis and extensive empirical exploration to find a suitable watermark level, weights size per transform, appropriate coefficients, and scaling criteria. Our studies show that DeepiSign can hide a 2-bits message in each coefficient, resulting in a total of 1KB secret that can be hidden in each layer without significantly impacting the model's accuracy.

\section{DeepiSign Methodology}\label{sec:methodology}

DeepiSign consists of two stages: (1) embedding the secret (before CNN deployment); and (2) retrieving the secret for verification (after CNN deployment). In the embedding stage ($\tilde{CNN}=f_e(s,h,CNN)$), a designed algorithm $f_e$ hides a pre-defined secret $s$ and a hash $h = H(s)$ where $H$ is a secure hash function\footnote{We use SHA256~(\url{https://docs.python.org/3/library/hashlib.html}).}. During the verification ($\tilde{s},\tilde{h}=f_r(\tilde{CNN})$), the algorithm $f_r$ retrieves a secret $\tilde{s}$ and its hash $\tilde{h}$ from $\tilde{CNN}$. $\tilde{s}$ and $\tilde{h}$ need to be further verified by calculating new $h_n$ from $\tilde{s}$. If $\tilde{h}$ and $h_n$ are the same, it confirms that the carrier data $\tilde{CNN}$ is pure and not changed. Otherwise, the carrier data is tampered by adversaries. The embedding algorithm to CNN models is summarized in Algorithm \ref{alg:ebeddingalgorithm}. \\

\begin{figure*}[!h]
	\centerline
	{\includegraphics[trim={0 1cm 0 2cm},clip,scale=0.25]{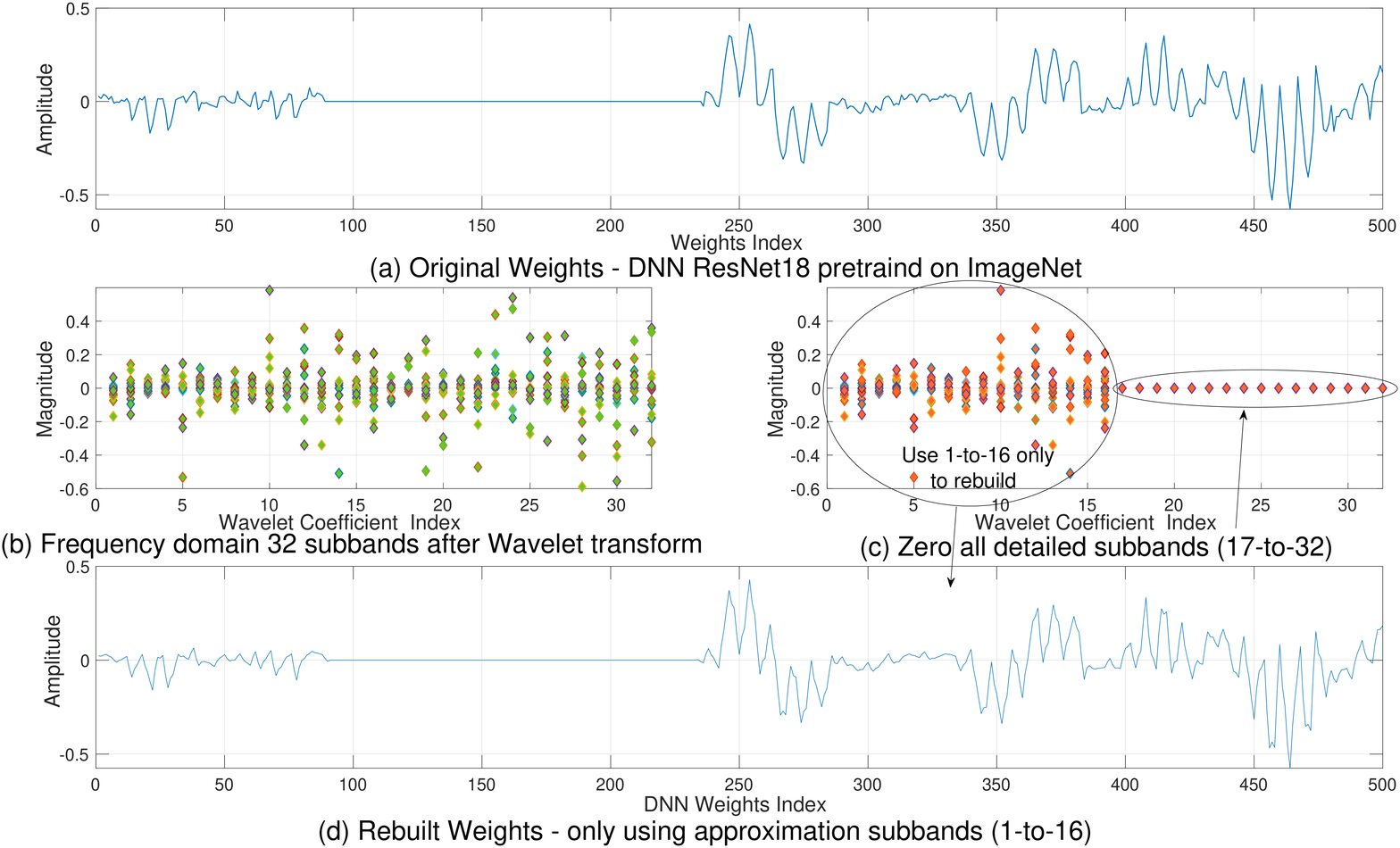}}
	\caption{Example of less important coefficients (17-to-32) in rebuilding the weights of ResNet18 model.}
	\label{fig:wipecoeff17_32}	
\end{figure*}
\begin{algorithm}[!h]
	\SetAlgoLined
	\textbf{Input}: CNN model\\
	\textbf{Output}: Protected CNN model\\
	$l_i$,$l_L$: $i$th and last layers of the model\\
	$\nu$: Scramble vector\\
	\nl $\nu \gets$ \textbf{Generate\_secret}(seed)\\
	\nl \For{$i\gets1$ \KwTo $l_L$}{
		\nl	$r \times c \gets$ \textbf{Reshape}($l_i$) // from 4D to 2D\\
		\nl	$M \times N \gets$ \textbf{Wavelet\_convert}($r \times c$)\\
		\nl	$s, h \gets$ \textbf{Prepare\_secret}($l_i$)\\
		\nl	$\tilde{s} \gets$ \textbf{Merge\_secret}($s,h$)\\
		\nl	$\tilde{M} \times \tilde{N}\gets$ \textbf{Generate\_scramble}($\nu$, $M \times N$)\\
		\nl	$\delta , \varrho \gets$ \textbf{Derive}($M \times N$)\\
		\nl	${M}'' \times {N}'' \gets$ \textbf{Scale}($M \times N$,$\delta , \varrho$)\\
		\nl	${M}'' \times {N}'' \gets$ \textbf{Hide}(${M}'' \times {N}''$, $\tilde{M} \times \tilde{N}$, $\tilde{s}$ )\\
		\nl	$M \times N \gets$ \textbf{Rescale}(${M}'' \times {N}''$, $\delta$ , $\varrho$)\\
		\nl	$r \times c  \gets$ \textbf{Wavelet\_inverse}($M \times N$)\\
		\nl	$l_i \gets$ \textbf{Shape}($r \times c$)  // from 2D to 4D
	}   	
	\caption{Embedding an Invisible Fragile Watermark}
	\label{alg:ebeddingalgorithm}
\end{algorithm}

\textbf{Summary of Algorithm \ref{alg:ebeddingalgorithm}}: we first reshape the hidden layer \textit{weights} (Section \ref{subsec:coefscramble}) to be ready for wavelet transform as depicted in line 3 (\textbf{Reshape}).  We then convert the \textit{weights} from the spatial domain to the frequency domain using the wavelet transform (Section \ref{sub:wavelet}) as shown in line 4 (\textbf{Wavelet\_convert}).  We next prepare a unique secret for each hidden layer, calculate its hash and merge them (Section \ref{subsec:key}) as appears in lines 5-6 (\textbf{Prepare/Merge}). We then generate a random matrix to ensure randomization of the hiding process (Section \ref{subsec:randomorder}) as shown in line 7 (\textbf{Generate\_scramble}). We scale the resultant coefficients (Section \ref{subsubsec:scale}) before hiding to preserve the sign and decimal accuracy as appears in lines 8-9 (\textbf{Derive/Scale}). We then start hiding the  secret bit-by-bit randomly in less significant coefficients following the random matrix (Section \ref{sub:hiding}) as shown in line 10 (\textbf{Hide}).  Functions in lines 11-13 are basically to rescale the coefficients, convert them back from the frequency domain to the spatial domain and finally shape the \textit{weights} back to their 4D form (Section \ref{sub:inverse}).  

\subsection{Reshaping Weights}\label{subsec:coefscramble}

We firstly pre-process the \textit{weights} in each CNN hidden layer $l_i$ so that they can be used for wavelet transform in the next stage. To achieve this, we reshape\footnote{We use a general reshapre function (\url{https://www.w3schools.com/python/numpy_array_reshape.asp}).} them from 4D ($a\times b\times c \times d$) to 2D ($r\times c$) form. For example, the weights in a ResNet18 hidden $l_{13}$ is reshaped from 4D (3x3x256x512) into 2D (4608 x 256).

\subsection{Converting CNN Weights to Frequency Domain}\label{sub:wavelet}

Hiding the secret directly into hidden layer weights may yield high distortion, leading to the degrading of the model accuracy. To solve this challenge, we employ Discrete Wavelet Transform (DWT) to convert the weights from their spatial domain into the frequency domain so that the most significant coefficients are preserved to rebuild the weights after hiding. Fig. \ref{fig:wipecoeff17_32} shows this process for CNN ResNet18: (a) the original block of weights, (b) converting the weights into the frequency domain using DWT, (c)  wiping out all detailed sub-bands as zeros (50\% of all) while maintaining the approximation sub-bands, and (d) converting back to the spatial domain for rebuilding the original \textit{weights} from the approximation sub-bands alone. The observations of the results from these steps motivate us to use the signal processing techniques to increase the capacity of hiding the secret with little distortion effect into the original model.

\begin{figure}[!ht]
	\centerline
	{\includegraphics[scale=0.13]{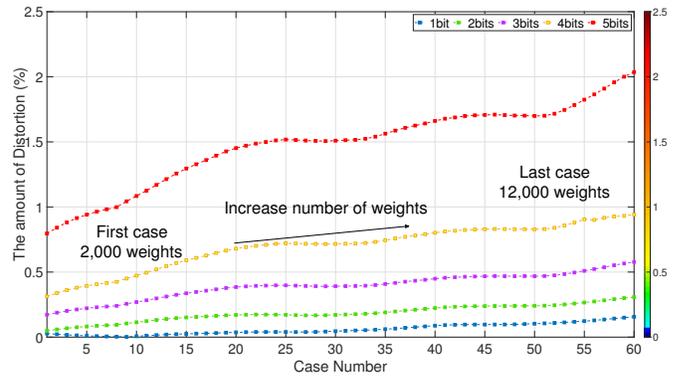}}
	\caption{Resultant distortion impacting CNN hidden layers from various watermark levels and the number of \textit{weights} per transform.}
	\label{fig:distortion}	
\end{figure} 

In our approach, we apply five levels of wavelet packet decomposition to each layer of a CNN model (e.g., ResNet18), which results in 32 sub-bands. A wavelet family, called Daubechies with the order 2 ($db2$), is chosen in the transformation process because its performance in analyzing discontinuous-disturbance-dynamic signals has already been proven in \cite{abuadbba2015wavelet,wavcomp:ning2011wavelet}. To minimize the distortion of the model, we do not change the low approximation sub-bands (i.e., from 1 to 16) because they represent the most significant features to rebuild the CNN layer's \textit{weights}. On the other hand, several bits are manipulated in the rest of the detailed sub-bands to embed secret bits; the number of bits that can be embedded is called the watermark level. Several experiments were performed to select an appropriate watermark level and the number of \textit{weights} per transform. As shown in Fig. \ref{fig:distortion}, we experimentally find that embedding two bits at all high-frequency sub-band coefficients results in a reasonable low distortion $\le 0.25\%$. We also find that using a large number of weights per transform may result in higher distortion. Accordingly, we keep the number of \textit{weights} per transform $\le 12000$ in all experiments. Note that our benchmark for the acceptable distortion is to maintain the accuracy of the original model.

\begin{figure*}[hbt!] 
	\centerline
	{\includegraphics[scale=0.55]{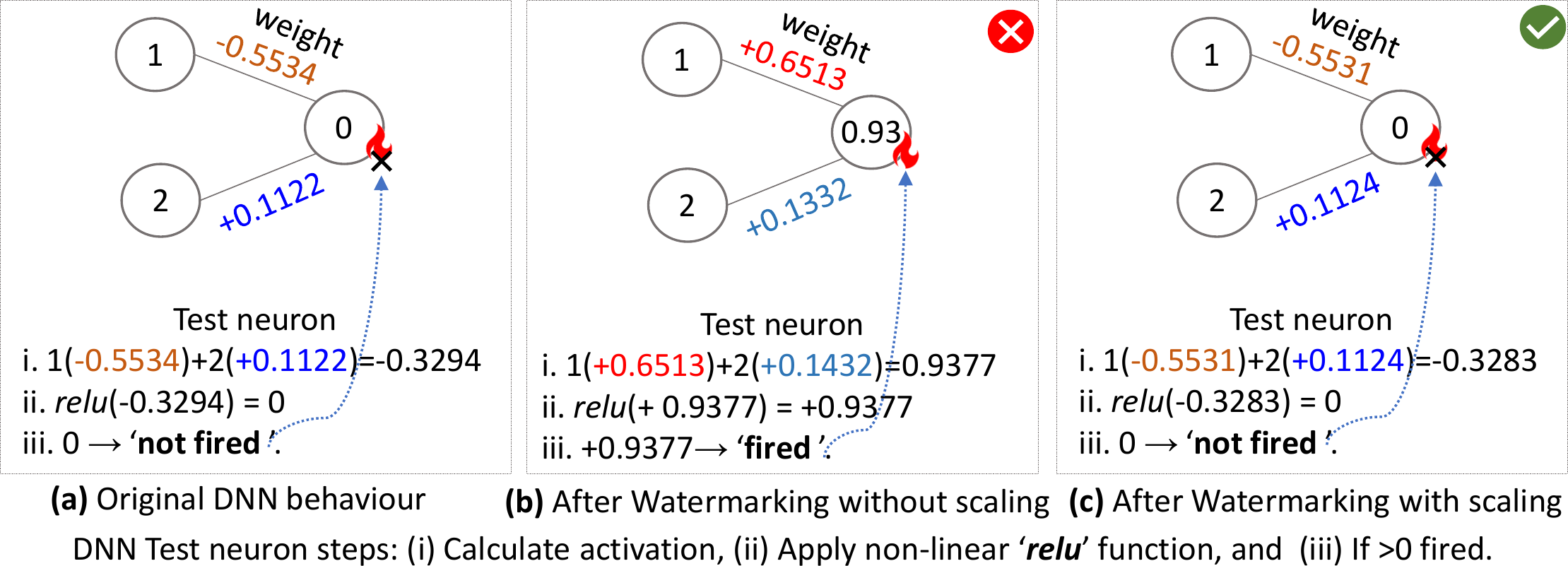}}
	\caption{Example of the impact of applying the derived $\delta$ and $\varrho$ before the hiding process. Not applying them may result in flipping the neurons activation as in (b) which leads to misclassification.}
	\label{fig:neurontest}
	\hfil
\end{figure*}
\subsection{Protecting the Embedded Secret}\label{subsec:stegoconfidentiality}

In watermarking, the secret can be exposed to attackers if the secret is always hidden at a fixed position. To embed the secret randomly, we use  a scrambling vector $\nu_{i\in[1,256]}$ pre-filled with random values. We assume that these this parameter is known only to authorized validators.

\textbf{Hashing the Secret: }\label{subsec:key}
Our secret includes: (1) structural information $s$ (data attributes as an arbitrary strings to stamp the model) and (2) the hash $h$ of the structural information $s$ using a secure hash function.
We then merge $s$ and $h$ on a bit level as shown in Eq. \eqref{eq:xor}.

\begin{equation} \label{eq:xor}
\widetilde{s}\Leftarrow \xi(s,h)
\end{equation}
where $\xi$ is a merging algorithm;  $s$ is the model secret; $h$ is its hash and $\widetilde{s}$ is the merged secret.

\textbf{Generating Scramble: }\label{subsec:randomorder}
To embed the merged secret $\widetilde{s}$ into randomly selected locations within CNN layers, we use the scrambling vector $\nu$ to create a random sequence of coefficients in the form of $2D$ matrix $Z$ (see Eq. \eqref{eq:selctedcoef}).

\begin{equation}\label{eq:selctedcoef}
Z\Leftarrow \begin{cases}
\tilde{M}= f_x(\nu)\\ 
\tilde{N}= {f_x}'(\nu) 
\end{cases}
\end{equation}
where $\tilde{M}$ and $\tilde{N}$  are the generated sequence of numbers; $f_x$ and  ${f_x}'$ are the scrambling functions. The combination of $\tilde{M}$ and $\tilde{N}$ is used to build a 2D $\tilde{M}\times \tilde{N}$ matrix $Z$ (see Eq. \ref{eq:smatrix}).
\begin{equation}\label{eq:smatrix}
Z\{\tilde{M},\tilde{N}\} =
\begin{bmatrix}
\tilde{m}_1,\tilde{n}_1 & \tilde{m}_1,\tilde{n}_2 & \cdots & \tilde{m}_1,\tilde{n}_n\\
\tilde{m}_2,\tilde{n}_1 & \tilde{m}_2,\tilde{n}_2 & \cdots & \tilde{m}_2,\tilde{n}_n\\
\vdots  & \vdots  & \ddots & \vdots \\
\tilde{m}_{\tilde{M}},\tilde{n}_1 & \tilde{m}_{\tilde{M}},\tilde{n}_2 & \cdots & \tilde{m}_{\tilde{M}},\tilde{n}_{\tilde{N}}
\end{bmatrix}
\end{equation}


\subsection{Scaling Coefficients}\label{subsubsec:scale}

To protect the accuracy of neurons at the hidden layers and preserve the sign of \textit{weights}, two factors are derived after analyzing millions of \textit{weights}. The first factor $\delta$ is used to ensure that all values are positive (e.g., the lowest value $+ (-1)$). The second factor $\varrho$ is used to maintain all four decimal values (e.g., $\times$ 10000) (see the impact in Fig. \ref{fig:neurontest}). $\delta$ and $\varrho$ are used to scale the coefficients before the embedding process so that the behavior of neurons in the networks is preserved.

\subsection{Embedding the Secret as Distributed Bits}\label{sub:hiding}

The  secret bits $\widetilde{s}$ are  embedded bit-by-bit in the scaled coefficients ${M}'' \times {N}''$ corresponding to $\tilde{M} \times \tilde{N}$ generated in the random order. $\tilde{M} \times \tilde{N}$ consists of pairs of random values to refer to positions of ${M}'' \times {N}''$. For $i^{th}$ two bits in the secret $\widetilde{s}$, we choose a scaled coefficient located at ($x_i$, $y_i$) in that matrix using the $i^{th}$ entry ($x_i$, $y_i$) of the scrambling matrix and replace the two least significant bits of the chosen coefficient with the two secret bits. 


\subsection{Inversing from Frequency Domain} \label{sub:inverse}

The resultant detailed coefficients after the hiding process are called \emph{marked} coefficients. At this stage, the \emph{marked} coefficients are rescaled and re-embedded back into the 32 sub-bands coefficients matrix before applying the inverse DWT to convert \textit{weights} from their frequency domain to their original space domain. The result of the reconstructed weights is called \emph{marked} \textit{weights} (containing the hidden secret), which are almost similar to the original \textit{weights}. The advantage of this approach is that the \emph{marked} \textit{weights} can be used for the prediction. However, only authorized validator (i.e., with $\kappa$ and $\nu$) can extract the secret and verify it. The inverse DWT is defined by Eq. \ref{eq:IDWT},
\begin{equation}\label{eq:IDWT}
X=\sum_{a}\sum_{b}Y(a,b)\Phi_{ab}(n)
\end{equation}
where $X$ is the \textit{weights} in their original time domain. Finally, the \textit{weights} are reshaped back from the 2D into their original 4D shape before integrating them into the CNN layer $l_i$. 

\begin{figure}[!ht] 
	\centerline
	{\includegraphics[scale=0.26]{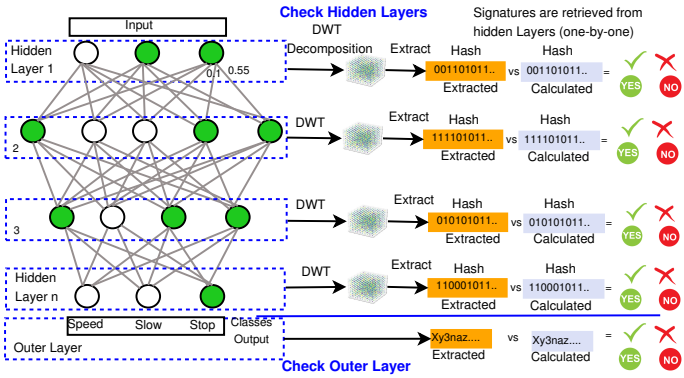}}
	\caption{Overall process for the watermark retrieval and validation.}
	\label{fig:validation}
	\hfil
\end{figure}

\subsection{Protecting All Hidden Layers} \label{subsec:layers}
The hiding steps explained in Algorithm \ref{alg:ebeddingalgorithm} are repeated for each hidden layer $l_{i\in[1,L]}$.  The steps of generating  the scramble matrix $\tilde M \times \tilde N$  are repeated for each hidden layer. The only difference between the layers is that we shift the index $i$ over $\nu$  by the hidden layer position $h$. Hence, we can generate a unique scrambling matrix for each layer. 

\subsection{Retrieving and Validating}

To accurately extract and validate the secret, Alice/Bob must have the scrambling vector $\nu$, and the protected model $\tilde{D}$. The process is nearly similar to the hiding steps, but the secret bits are recovered rather than embedded. Fig. \ref{fig:validation} demonstrates the required steps. First, \textit{weights} at each layer, $l_i$, are fetched and shaped  before applying DWT. Then, the detailed coefficients are selected and scaled. Next, the random hiding order is generated using $\nu$ and followed to retrieve the secret bits. Finally, we then calculate their hash, and verify it against the embedded hash. Thus, a slight change, even in one layer, can be detected and highlighted. 

\section{Experiments}

\textbf{Experiment Steps:} Our experiment steps can be summarised as follows: (1) Train ResNet18 architecture with  MNIST, CIFAR-10 and Imagenet training datasets; we name the resultant models  $D_1$, $D_2$ and $D_3$. (2) Evaluate the classification baseline accuracy of  $D_1$, $D_2$ and $D_3$ using MNIST, CIFAR-10 and Imagenet testing datasets. (3) Apply our \emph{watermark} technique ``DeepiSign'' to  $D_1$, $D_2$ and $D_3$; we name the obtained models $\tilde{D_1}$, $\tilde{D_2}$ and $\tilde{D_3}$. (4) Evaluate the classification accuracy of  $\tilde{D_1}$, $\tilde{D_2}$ and $\tilde{D_3}$ using MNIST, CIFAR-10, and Imagenet testing datasets. (5) Verify the accuracy of the retrieved information after the extraction process using the hash. (6) Perform manipulation attacks on $\tilde{D_1}$, $\tilde{D_2}$ and $\tilde{D_3}$ using various adversarial mechanisms and verify the integrity of the hidden bits. (7) We test DeepiSign on 3 other widely-used architectures (i.e., AlexNet, VGG16 and MobileNetV2) to ensure its effectiveness. (8) We finally perform an out of the box experiment to validate the accuracy impact using randomly picked samples from the Internet.  

\textbf{Configuration:} To obtain the neutral and unbiased results, all experiments were performed using  $|\nu|=256$. \textit{weights} size per transform varies from 4000 to 12000 and a maximum of 2-bits are hidden in every selected detailed coefficient (see Fig. \ref{fig:distortion}). A total of 8,192 bits (i.e., around 1KB) is embedded in each hidden layer. All deep layers have been protected with our technique. We optimized all models using Stochastic Gradient Descent (SGD) and an initial learning rate of 0.0001. We also used 10 epochs with a batch size of 100, and factor 10 for both \textit{weights} and \textit{bias} learning rates.

In the following, we investigate whether DeepiSign can satisfy the three conditions discussed in Section~\ref{sec:intro}.\\

{\bf(C1) Can DeepiSign Maintain the Accuracy?}

\begin{figure}[!h]
	\centerline
	{\includegraphics[scale=0.43]{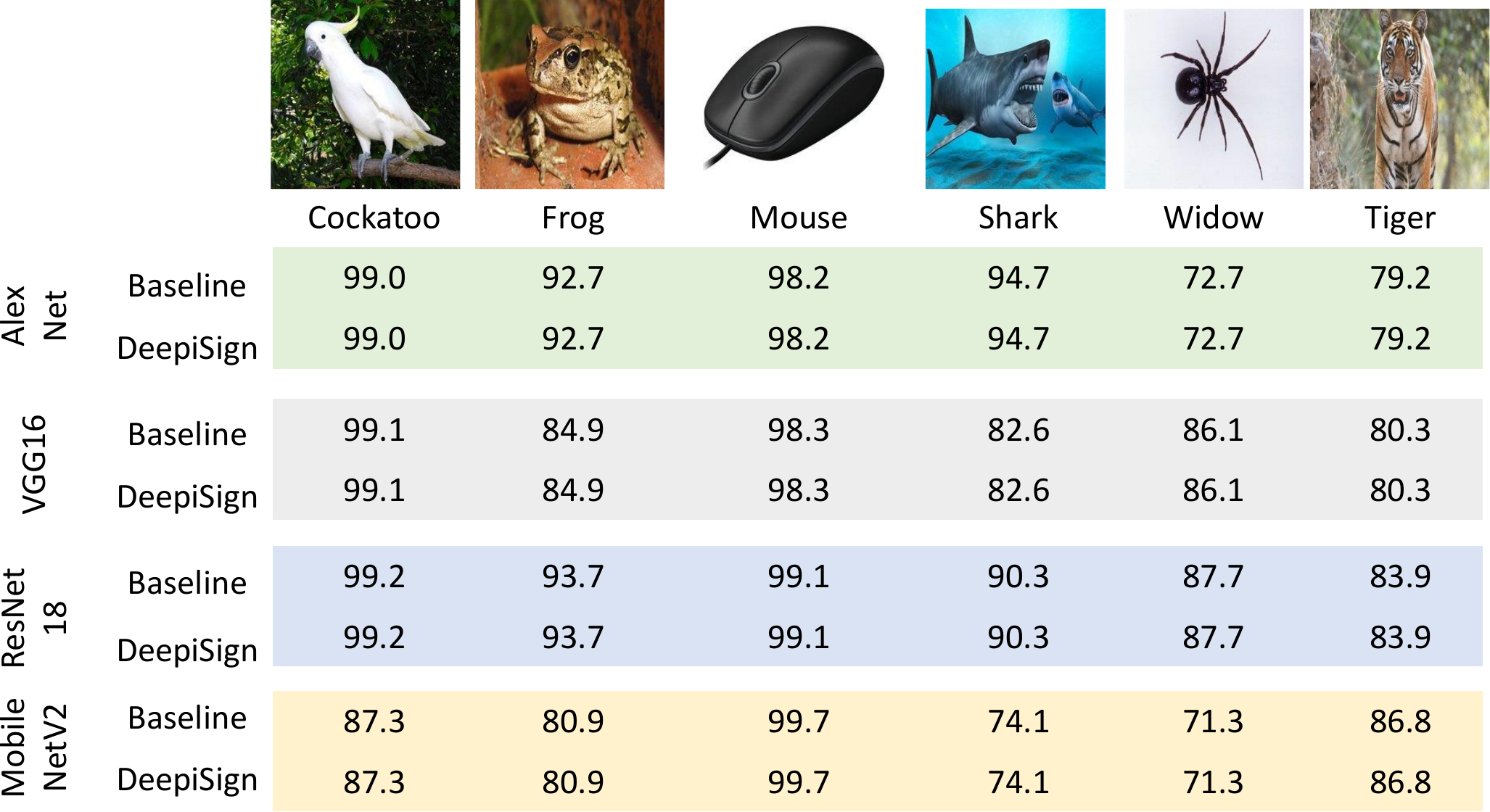}}
	\caption{Classification accuracy results using sample images before and after applying DeepiSign on AlexNet, VGG16, ResNet18 and MobileNetV2.}
	\label{fig:extra_images_7}
\end{figure}

Table \ref{tb:accuracy} shows the classification accuracy of the three trained models obtained from the three datasets (MNIST \cite{ds:mnist:le1995}, CIFAR-10 \cite{ds:cifar10:kriz2009}, and Imagenet \cite{ds:deng2009imagenet}) before and after applying our DeepiSign technique with different watermark levels. The experiments match with our early observations shown in Fig. \ref{fig:distortion}. Once we exceed the watermark level of 2-bits per coefficient, the accuracy starts to be affected. We observe that changing more than 2 bits in the frequency domain coefficients results in flipping the first decimal values of the rebuilt \textit{weights}--such flip impacts on the set of neurons to be activated. On the contrary, the watermark level of 2 bits affects only the fourth decimal values and rarely the third decimals of the detail coefficients. Hence, it has little effect on the two significant factors (i.e., the sign and first decimal) of the rebuilt \textit{weights}. Bit Error Rate (BER) is 0\%, which means the secrets are verifiable with no errors. We further evaluate the accuracy of VGG16, AlexNet and MobileNetV2 before and after applying the best watermark level of 2-bits. The obtained accuracy results are the same as AlexNet 79.51\%, VGG16 81.80\%, and MobileNetV2 74.25\%. BER is 0\% in both cases, indicating that DeepiSign has no noticeable impact on the models' accuracy. Fig. \ref{fig:extra_images_7} presents an out of the box experiments with random samples. 

\textbf{Summary:} The answer is affirmative, where DeepiSign can maintain the accuracy by embedding 2 bits in the less significant frequency domain coefficients.

\begin{table}[!ht]
	\caption{Comparison between baseline classification accuracy (\%) before and after applying our DeepiSign with different watermark levels.}
	\label{tb:accuracy}
	\centering
	\begin{tabular}{|l|c|c|l| l|l|c|}\hline
		\multicolumn{2}{|c|}{} & \multicolumn{4}{c|}{DeepiSign -  \emph{watermark} levels}  &  \\\hline 
		Dataset  & \multicolumn{1}{|l|}{\textbf{Baseline}} & \multicolumn{1}{l|}{1-bit} & \textbf{2-bits}  & 3-bits  & 4-bits  & BER \\\hline 
		MNIST    & \textbf{99.00}                               & 99.00                              & \textbf{99.00} & 98.97 & 98.97 & 0   \\\hline
		CIFAR-10  & \textbf{88.40}                               & 88.40                              & \textbf{88.40} & 88.38 & 88.38 & 0   \\\hline
		Imagenet & \textbf{67.71}                               & 67.71                              & \textbf{67.71} & 67.63 & 67.63 & 0\\  \hline
	\end{tabular}
\end{table}
\begin{table*}[!ht]  
	\caption{Integrity verification results of the marked and the poisoned models. (\checked) means hidden secrets and its hash match correctly with 0\% BER. ($\times$) means hidden secrets and its hash mismatch with error \% shown in BER.}
	\label{tb:inputpoison}
	\centering
	\begin{tabular}{| c | c | c | c | c | c | c |}
		\hline 
		&    \multicolumn{2}{|c|}{MNIST}  &  \multicolumn{2}{|c|}{CIFAR-10}  &  \multicolumn{2}{|c|}{Imagenet}\\ \hline 
		\pbox{8cm}{Test}  &\pbox{17cm}{Marked $\tilde{D_1}$}  & \pbox{17cm}{Poison   $\tilde{D^{adv}_1}$}  & \pbox{17cm}{Marked $\tilde{D_2}$}  &\pbox{17cm}{ Poison   $\tilde{D^{adv}_2}$}  & \pbox{17cm}{Marked $\tilde{D_3}$}  & \pbox{17cm}{Poison   $\tilde{D^{adv}_3}$} \\ \hline 				
		BER &	0\%	&48-61\%	&0\%	&52-61\% &0\%	&46-63\%\\ \hline
		\pbox{8cm}{ResNet18} &\checked	&$\times$	&\checked	&$\times$ &\checked	&$\times$\\ \hline
		\pbox{8cm}{VGG16} &\checked	&$\times$		&\checked	&$\times$ &\checked	&$\times$\\ \hline
		\pbox{8cm}{AlexNet} &\checked	&$\times$		&\checked	&$\times$ &\checked	&$\times$\\ \hline
		\pbox{8cm}{MobileNet} &\checked	&$\times$		&\checked	&$\times$ &\checked	&$\times$\\ \hline
		
	\end{tabular}
\end{table*}
{\bf(C2) Can DeepiSign Detect Model Manipulation Attacks?}

We consider two legitimate parties involved: CNN model provider (Alice) and CNN model customer (Bob). Alice trains a CNN locally or in a secure location. We assume that Alice and Bob are honest. 
We consider a problem similar to \cite{rw:he2018verideep}, where the customer Bob wants to verify the integrity and authenticity of the CNN model. However, in our case, this should be done locally (i.e., to avoid a single point of failure at a remote site) and automatically at the customer end. Similarly, CNN model provider Alice wants to verify the model in addition to checking the dataset used and parameters in case of any dispute. We consider that an adversary (Eve) obtains the access to the CNN model similar to attacks demonstrated in \cite{attacks:gu2017badnets,rw:liu2017trojaning} and performs one of the following attacks to compromise the integrity of the model either during the deployment or at  Bob's end.

\begin{figure}[!t]
	\centerline
	{\includegraphics[scale=0.65]{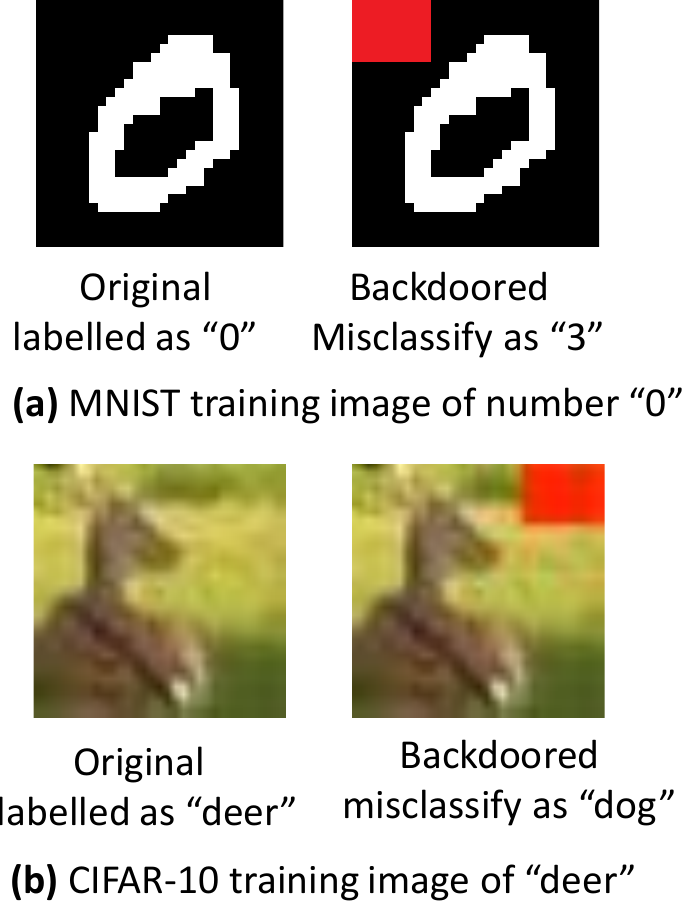}}
	\caption{Two random images from MNIST and CIFAR-10 before and after inserting a backdoor (i.e., red sticky note).}
	\label{fig:poisonimage}	
\end{figure}

\textbf{Attack 1 - Targeted Input Poisoning: } It is an attack in which the attacker retrains the model by using at least one sample and corresponding label (not reflecting the ground truth) \cite{attacks:gu2017badnets}.  We implement this attack on  MNIST,  CIFAR-10 and Imagenet protected models by generating $\tilde{D_1}$, $\tilde{D_2}$ and $\tilde{D_3}$.  Fig. \ref{fig:poisonimage} shows two examples: (1) we randomly pick $g_1$ from MNIST as number `0', insert a backdoor, and reinject it as number `3'. In other words, $\tilde{D_1}$ misclassifies number `0' as number `3' only when seeing an image of digit `0' with a backdoor. (2) We also randomly pick $g_2$ from CIFAR-10 as an image of a `deer', insert a backdoor, and reinject it as a `dog'. We retrain the protected $\tilde{D_1}$ and $\tilde{D_2}$ with the poisoned MNIST and CIFAR-10 and obtained $\tilde{D^{adv}_1}$ and $\tilde{D^{adv}_2}$. We performed  similar steps on Imagenet and obtained $\tilde{D^{adv}_3}$. All training parameters are the same. Finally, the accuracy and integrity of all hidden layers, the hidden hash, BER of $\tilde{D^{adv}_1}$, $\tilde{D^{adv}_2}$ and $\tilde{D^{adv}_3}$ are carefully examined and presented in Table \ref{tb:inputpoison}. 

\textbf{Findings:} From the accuracy perspective, it is clear that there is no degradation; this means such attacks are hard to detect by observing the accuracy. However, our DeepiSign algorithm was able to detect this attack in all cases.

\textbf{Attack 2 - Output Poisoning:} In this type of attack, only the output classes are tampered to cause misclassification \cite{rw:adi2018turning}. We implement this attack by only manipulating one class in the output layer of MNIST, CIFAR-10, and Imagenet protected models. We flip the class `0' with `3' in MNIST, the class `deer' with `dog' in CIFAR-10, and the class `mouse' with `keyboard' in Imagenet. 

\begin{figure}[!t]	
	\centering{\includegraphics[scale=0.5]{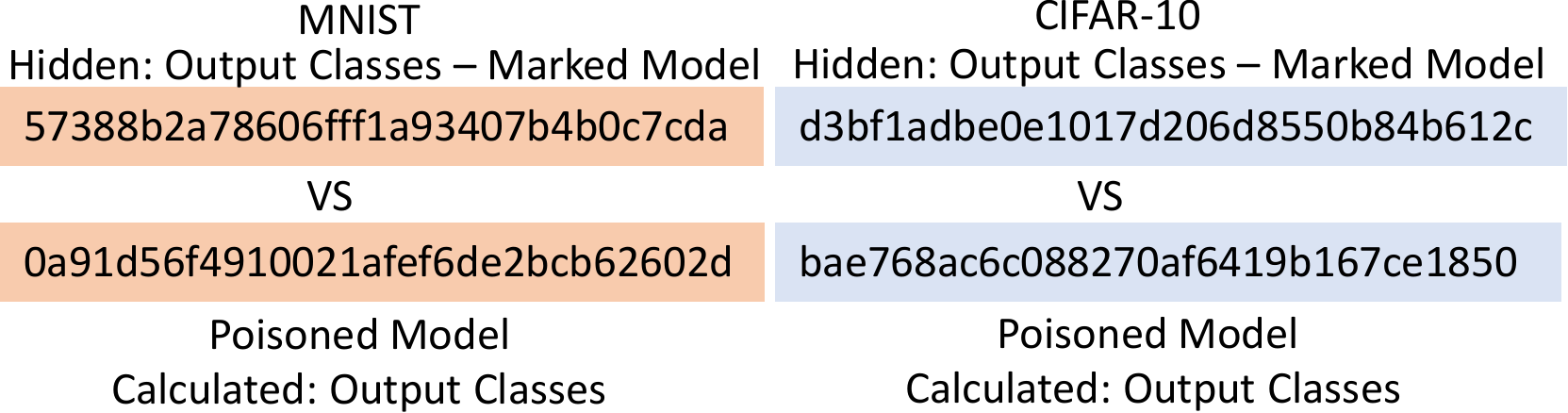}}
	\caption{Comparison between the embedded hash value and the hash value calculated from the poisoned output layer.}
	\label{fig:OutputPoison}	
\end{figure}

\textbf{Findings:} Although all deep layers are the same, DeepiSign can detect the attack (see Fig. \ref{fig:OutputPoison}) because the hash value embedded in each hidden layer is different from the hash value calculated from the poisoned output layer.  

\textbf{Attack 3 - Fine-tuning: } It is another type of attack that an adversary Eve uses to slightly manipulate the model, which may degrade or even improve the accuracy \cite{wm:ro2018deepsigns}. We implement this attack by only changing one parameter, which is the learning rate from (\textit{0.0001}) to (\textit{0.001}). The main reason for choosing the learning rate is that we do not want to manipulate many parameters to induce bias and make the attacks easily detectable. Table \ref{tb:finetune}  presents a result comparing the protected model with the fine-tuned model.

\begin{table*}[!ht]  
	\caption{Integrity verification results of the marked and the fine-tuned models.}
	\label{tb:finetune}
	\centering
	\begin{tabular}{| c | c | c | c | c | c | c |}
		\hline 
		&    \multicolumn{2}{|c|}{MNIST}  &  \multicolumn{2}{|c|}{CIFAR-10}  &  \multicolumn{2}{|c|}{Imagenet}\\ \hline 
		\pbox{8cm}{Test}  &\pbox{17cm}{Marked $\tilde{D_1}$}  & \pbox{17cm}{Poison   $\tilde{D^{adv}_1}$}  & \pbox{17cm}{Marked $\tilde{D_2}$}  &\pbox{17cm}{ Poison   $\tilde{D^{adv}_2}$}  & \pbox{17cm}{Marked $\tilde{D_3}$}  & \pbox{17cm}{Poison   $\tilde{D^{adv}_3}$} \\ \hline 				
		BER &	0\%	&54-65\%	&0\%	&57-71\% &0\%	&61-71\%\\ \hline
		\pbox{8cm}{ResNet18} &\checked	&$\times$	&\checked	&$\times$ &\checked	&$\times$\\ \hline
		\pbox{8cm}{VGG16} &\checked	&$\times$		&\checked	&$\times$ &\checked	&$\times$\\ \hline
		\pbox{8cm}{AlexNet} &\checked	&$\times$		&\checked	&$\times$ &\checked	&$\times$\\ \hline
		\pbox{8cm}{MobileNet} &\checked	&$\times$		&\checked	&$\times$ &\checked	&$\times$\\ \hline
		
	\end{tabular}
\end{table*} 

\textbf{Findings:} Despite a slight increase in the accuracy, our DeepiSign technique can detect the attack from both BER and the hash of the hidden layers. 

\textbf{Summary:} DeepiSign satisfies the second condition by detecting 3 manipulation attacks on 3 CNN architectures. \\

\textbf{(C3) Can DeepiSign Provide Sufficient Security?}

In the DeepiSign design, we focus on the threat model where an adversary (Eve) has access to the protected CNN model; Eve's task of detecting the embedded secret is steganalysis. Steganalysis has been widely studied in the multimedia domain (e.g., Image, video, and audio) \cite{steganalysis:17:audio}. The steganalysis in the multimedia domain is designed to find abnormal patterns among neighbor pixels to detect the steganography or invisible watermark \cite{steganalysis:14}. Only a few studies attempted to apply fragile watermarking to non-multimedia data such as time-series data (e.g., ECG and sensor streams) where there is no known correlation between adjacent data values \cite{prd:results2013,abuadbba2016walsh}. In this paper, we follow a theoretical steganalysis of non-multimedia watermarking suggested in \cite{prd:results2013} in terms of confidentiality, integrity, and authenticity.

\textbf{Confidentiality Strength:} It is achieved with the scramble vector $\nu$. Assume $|\nu| \geq256$. Then, an attacker has to search $2^{256}$ to find $\nu$, which yields a 256-bit level strength.


\textbf{Integrity and Authenticity Strength:} To guarantee the authenticity and prevent retrieving the hidden information, the 32 sub-bands coefficients matrix after wavelet decomposition of \textit{weights} should have a suitable size (e.g., $\geq 4000$) as in $T=\sum_{i=1}^{r}R! \times \sum_{j=t}^{c}C!$. Where $T$ is the total number of possibilities; $R$ and $C$ are the rows and columns, respectively, of the 32 sub-bands coefficients matrix; and $t$ is the selected detailed coefficients that can be used from each row. Assume 4096 \textit{weights} from only one layer $l_3$, and their 32 sub-bands coefficients are in the size of  $128 \times 32$ after applying wavelet. If we assume that the threshold $t$ is $16$, $T$ can be calculated as $T=\sum_{i=1}^{128}128! \times \sum_{j=16}^{32}32!  \Rightarrow T\cong 8.068256\times 10^{194}$. 

\textbf{Summary:} We can see that it is computationally infeasible to break DeepiSign confidentiality, integrity, and authenticity in a reasonable time.

 \section{Related Work}\label{sec:relatedwork}

This section provides a brief review of related work on the attacks and defenses on CNN model integrity.

{\textbf{Poisoning attacks:}} Several techniques have been proposed in the literature to violate CNN integrity by inserting backdoors. Gu et al. \cite{attacks:gu2017badnets} introduced a poisoning attack in their BadNets work. They generated a poisoned model by retraining the original one with a poisoned training dataset. The attacked model behaves almost like the benign one except when the backdoor sign is encountered. They also showed that the backdoor remains active even after the transfer learning (to a new model). Liu et al. \cite{rw:liu2017trojaning} further improved this attack by tampering only a subset of weights to inject a backdoor. Chen et al. \cite{rw:chen2017targeted} proposed another attack where the attacker does not need to have access to the model. 

{\textbf{Poisoning Defenses:}} Defense against backdoor attacks is an active research area of research. Liu et al. \cite{rw:liu2017neural} introduced three different defense mechanisms: (1) Employing anomaly detection in the training data: such a method requires access to the poisoned dataset, which is unrealistic in practice; (2) Retraining the model to remove the backdoors -- however, retraining does not guarantee the complete removal of backdoors as demonstrated by previous work in \cite{attacks:gu2017badnets}; (3) Preprocessing the input data to remove the trigger -- it needs the adversary's aids, which is hard to achieve. Liu et al. \cite{rw:liu2017trojaning}  suggested that detecting the backdoor might be possible by analyzing the distribution of mislabelled data. However, the victim needs to feed the model with a large dataset of samples, rendering such an approach inefficient and expensive. He et al. \cite{rw:he2018verideep} recently introduced a defense technique by generating sensitive input samples to spot possible changes in hidden weights and produce different outputs. Bolun et al. \cite{wang2019neural} also demonstrated mitigation techniques using input filters, neuron pruning and unlearning to identify backdoors.
However, these defense techniques lack a mechanism to provide the integrity and authenticity of the hidden and outer layers.
This stream of work is very promising in a black-box setup to determine if the incoming input is benign or adversarial. However, these techniques cannot still determine if poisoning attacks compromise a CNN model.  


{\textbf{Watermarking: }} Several proposals were made to use watermarking to protect the Intellectual Property (IP) of CNN models. Uchide et al. \cite{rw:uchida2017embedding,rw:nagai2018digital} proposed a method of embedding a small watermark into deep layers to protect the owner's IP. This work provides a significant leap as the first attempt to watermark neural networks. Zang et al. \cite{rw:zhang2018protecting} further extended the technique to the black-box scenario. Merrer et al. \cite{rw:merrer2017adversarial} introduced 1-bit watermark that is built upon model boundaries modification and the use of random adversarial samples that lie near the decision boundaries. Rouhani et al. \cite{rw:rouhani2018deepsigns} proposed an IP protection watermarking technique that not only protects the static weights like previous works but also the dynamic activations. Recently, Adi et al. \cite{rw:adi2018turning} extended the backdoor approach into a watermarking scheme by inserting a backdoor to claim the ownership of the model. However, these studies have only focused on the ownership of the models by building a persistent watermark. When a model is poisoned or fine-tuned, watermarks should remain the same to ascertain the ownership. To the best of our knowledge, we are not aware of previous attempts that use fragile invisible watermarking to protect the integrity and authenticity of CNN models. 

\section{Conclusion}\label{sec:conclusion}

We propose a novel self-contained invisible mechanism, called DeepiSign, to protect CNN models' integrity and authenticity. DeepiSign embeds a secret and its hash into a CNN model securely to provide the model's integrity and authenticity. To reduce the distortion due to hiding, which is inherent to watermarking, DeepiSign uses a wavelet-based technique to transform each layer's weights from the spatial domain to the frequency domain. To preserve accuracy, it utilizes the less significant coefficients to hide the secret using both secure key and scramble vector. We performed theoretical analysis as well as empirical studies. The analysis showed that DeepiSign could hide about 1KB secret in each layer without degrading the model's accuracy. Several experiments were performed on three pre-trained models using three datasets against three types of manipulation attacks. The results prove that DeepiSign is verifiable at all times with no noticeable effect on classification accuracy, and robust against a multitude of known CNN manipulation attacks.

\section*{Acknowledgment}
The work has been supported by the Cyber Security Research Centre Limited whose activities are partially funded by the Australian Government’s Cooperative Research Centres Programme. This work was also supported by the National Research Foundation of Korea (NRF) grant funded by the Korea government (MSIT) (No. 2019R1C1C1007118).
\bibliographystyle{ACM-Reference-Format}
\bibliography{References}

\end{document}